\documentclass[12pt,preprint]{aastex}
\usepackage{natbib}

\begin{document}

\title{Problems in suppressing cooling flows in clusters of galaxies
by global heat conduction}

\author{Noam Soker\altaffilmark{1}}
\altaffiltext{1}{ Department of Physics, Oranim, Tivon 36006, Israel;
soker@physics.technion.ac.il.}

\begin{abstract}
I use a simple analytical model to show that simple heat conduction
models cannot significantly suppress cluster cooling flows.
I build a static medium where heat conduction globally balances
radiative cooling, and then perturb it.
I show that a perturbation extending over a large fraction of the
cooling flow region and with an amplitude of $\sim 10 \%$,
will grow to the non-linear regime within a Hubble time.
Such perturbations are reasonable in clusters which frequently
experience mergers and/or AGN activity.
This result strengthens previous findings which show that a steady
solution does not exist for a constant heat conduction coefficient.
\end{abstract}

{\bf Key words:} 
galaxies: clusters: general ---
cooling flows ---
intergalactic medium ---
X-rays: galaxies: clusters

$$
$$

\section{Introduction}

Recent {\it Chandra} and {\it XMM-Newton} observations resolved
the inner structure of the cooling flow region in
clusters of galaxies.
Traditionally, this region is defined as the region where radiative
cooling time is shorter than the age of the cluster;
not all clusters possess cooling flows (e.g., Fabian 1994).
While hot, $T \simeq 3-8 \times 10^7$ K, X-ray emitting gas
seems to cool at rates of $10-500 \, M_\odot$ yr$^{-1}$, there
are almost no indications for gas cooling below a temperature of
$\sim 10^7$ K (e.g., Kaastra et al. 2001; Peterson et al.\ 2001, 2002;
Fabian 2002).
This renewed interest in an old idea (e.g., Bregman \& David 1988),
that heat conduction from the intra-cluster medium (ICM) outside
the cooling flow region supplies the energy carried by radiation
(e.g., Narayan \& Medvedev 2001; Voigt et al.\ 2002;
Fabian, Voigt, \& Morris 2002; Ruszkowski \& Begelman 2002).
Bregman \& David (1988) show that the heat conduction needs a fine
tuning in order to explain the temperature and density profiles
in cooling flows, and that a solution with a constant heat conduction
coefficient is unstable, i.e., no steady solution exist. 
Norman \& Meiksin (1996) try to increase the heat conduction
efficiency by allowing the cooling flow to stretch magnetic flux loops,
which then reconnect;
cooling flow is not completely suppressed, but the cooling rate
is substantially reduced in their model.
Other works argue, from different reasoning, for a heat conduction
coefficient $\kappa$ much bellow the Spitzer value for
unmagnetized plasma, such that heat conduction does not play a
global role in clusters and cannot prevent cooling flows
(e.g., Pistinner \& Shaviv 1996; Markevitch, Vikhlinin, \& Forman 2002;
Loeb 2002).

In a recent paper Zakamska \& Narayen (2003; hereafter ZN03)
fit five cooling flow clusters with a simple hydrostatic model
where heat conduction balances radiative cooling.
The thermal conductivity is $\sim 30 \%$ the Spitzer conductivity.
They argue that this model is stable against local transverse
perturbations.
For five cooling flow clusters they could not fit such a model, as they
require too high conductivity.
In this paper I show that such a solution is unstable to large
large-scale radial perturbation with amplitudes $\gtrsim 10 \%$ 
in the temperature, over the age of a cluster.
Hence, either the heat conduction does not play a role, or the
cooling flow is much younger than the cluster, as proposed
in the moderate cooling flow model (Soker et al. 2001). 
In either cases, heat conduction does not help in explaining
the large scale behavior in cooling flow clusters, although
it may be important locally, i.e., conduction to cold clouds
in the inner region.
In $\S 2$ I build a simple hydrostatic model with a power-law ICM.
In $\S 3$ I use this model to argue that the solution is not
stable over a long time, hence strengthening the results of
Bregman \& David (1988).
{{{ In light of the renewed interest in heat conduction,
(sometimes ignoring previous results), I aim at reaching this
conclusion via a simple model. }}}
My short summary is in $\S 4$.

\section{A Simple Steady State Hydrostatic Solution}

To show that a solution where conduction balances radiative cooling
is unstable, I start by building a simple spherically symmetric
model with a power-law medium
\begin{equation}
P=P_0 r^{-\eta}; \qquad
T=T_0 r^{\beta}; \qquad
n=n_0 r^{-\eta-\beta},
\label{eq:medium}
\end{equation}
where $P$, $T$, and $n$ are the thermal pressure, temperature,
and electron density, respectively, and $r$ is a dimensionless
radial coordinate.
{{{ The profiles used in equation (1) imply that the treatment
is correct for $r>0$, or more practically for $r \gtrsim 0.1$.
This does not change the conclusions of the present paper, because
the instability found here and most of the radiative energy loss
occur at much larger distances from the center. }}}
Like ZN03 I take the cooling rate per unit volume and the heat
flux to be $j=K_1 n^2 T^{1/2}$, and $F= - f K_2 T^{5/2} dT/dr$,
respectively, where $K_1$ and $K_2$ are constants, and $f$ is
the ratio of the conductivity to the Spitzer value.
The total radiated power, {{{{ a free-free emission,}}}}
inside a radius $r$ is given by
\begin{equation}
L_{\rm radS} = \int_0^r K_1 n^2 T^{1/2} 4 \pi x^2 d x
=8 \pi K_1 n_0^2 T_0^{1/2} (6-4 \eta -3 \beta)^{-1}
r^{3-2 \eta-\frac{3}{2}\beta}, 
\label{eq:lrad}
\end{equation}
where subscript `S' stands for the steady state solution.
Most of the energy is radiated just inward to $r$, such that the
detailed structure close to the center is not important.
The rate of inward heat flow across a spherical surface at $r$ is
\begin{equation}
L_{\rm heatS} = - 4 \pi r^2 F =
4 \pi f K_2 \beta T_0^{7/2} r^{\frac{7}{2} \beta +1}.
\label{eq:lheat}
\end{equation}
In a steady state hydrostatic model $L_{\rm heatS}=L_{\rm radS}$
for each radius, which implies two equalities:
the equality of the powers of $r$ in equations (\ref{eq:lrad})
and (\ref{eq:lheat}) and the equality of the constant terms in
these two equations.
The first equality gives
\begin{equation}
\eta=1-2.5 \beta,
\label{eq:eta}
\end{equation}
and the second equality gives
\begin{equation}
f K_2 \beta T_0^3
= 2 K_1 n_0^2 (6-4 \eta -3 \beta)^{-1}.
\label{eq:k1k2}
\end{equation}

From the solutions of ZN03, I find $\eta \sim 0.3$.
Equation (\ref{eq:k1k2}) implies that when $f$ is changed by
a factor $\alpha$, this may result in a modest change in
the temperature and a change in the density by a factor of
$\sim \alpha^{1/2}$; this is compatible with the finding of ZN03.

\section{Perturbing the Temperature Profile}
I take a simple form for the large scale temperature
radial perturbation, $\Delta T = T_\delta (r-1)$, such that the
new temperature profile is
\begin{equation}
T^\prime = T + \Delta T = T_0 r^{\beta}+ T_\delta (r-1),  
\label{eq:Tprime}
\end{equation}
where $T$ is given in equation (\ref{eq:medium}),
and the derivative of the temperature is {{{ (for $r>0$) }}}
\begin{equation}
\frac {d T^\prime}{dr} = \beta T_0 r^{\beta-1}+T_\delta.   
\label{eq:dTprime}
\end{equation}
For simplicity I assume that the pressure profile does not change,
hence the new density profile is
\begin{equation}
n^\prime = n (T/T^\prime). 
\label{eq:nprime}
\end{equation} 
{{{ A full accurate treatment should consider the perturbation in
pressure as well. The pressure perturbation is neglected here for the
following reasons. (1) As stated in Section 1, the aim of the present
paper is to present a simple treatment in order to clearly
demonstrate the global radiative instability.
(2) This assumption is accurate for a case of constant pressure.
Although the pressure in clusters is not constant, it is still quite
shallow. As derived below, the instability which occurs for a
constant pressure is not much different from the one for a shallow
pressure profile.
(3) Although being simple, the result of the
present treatment is in accord with the numerical results of
Bregman \& David (1988). This makes the treatment trustworthy. }}}

Only the first order in $T_\delta$ is retained. For example,
\begin{equation}
(T^\prime)^{5/2} =
[T_0 r^{\beta}+T_\delta (r-1)]^{5/2} =
T_0^{5/2} r^{5 \beta / 2} \left[ 1 + \frac{5}{2} \frac {T_\delta}{T_0}
r^{-\beta} (r -1) \right] + O(T_\delta^2).
\label{eq:T52}
\end{equation}
{{{ Keeping only the first order is not justified very close to the
center where the unperturbed temperature is very low.
As noted earlier, this does not affect the conclusions here, because
the instability occurs farther out in the cluster, and because in
real clusters the very central region is expected to be dominated
by the AGN activity any how. }}}
Substituting equations (\ref{eq:dTprime}) and (\ref{eq:T52}) in
the expression for the total rate of inward heat flow across a spherical
surface of radius $r$, similar to equation ({\ref{eq:lheat}) for
the steady state heat flow $L_{\rm heatS}$, gives
to first order in $T_\delta$
\begin{equation}
L_{\rm heat} = L_{\rm heatS}
\left[ 1+ \frac {T_\delta}{T_0 r^{\beta}} \left(
\frac{1}{\beta} r + \frac{5}{2} r - \frac{5}{2} \right) \right].
\label{eq:lh}
\end{equation}
Expanding $(n^{\prime})^2 (T^{\prime})^{1/2}$ to first order in
$T_\delta$, and integrating, similar to equation (\ref{eq:lrad})
for the steady state cooling rate $L_{\rm radS}$, and using
equation (\ref{eq:eta}) for $\eta$, gives
\begin{equation}
L_{\rm rad} = L_{\rm radS}
\left[ 1+ \frac {3}{2} \frac {T_\delta}{T_0 r^{\beta}} \left(
\frac{2+7 \beta}{2+5 \beta} - \frac{2+7\beta}{4+5\beta} r \right) \right].
\label{eq:lr}
\end{equation}

The net cooling rate of the region inward to $r$ is given by
$L_{\rm rad} - L_{\rm heat}$, which result in a cooling time
$\tau_{\rm cool}(r)$.
Let $\tau_{\rm rad}(r)$ be the cooling time calculated when only
the radiative cooling $L_{\rm radS}$ is considered, i.e.,
the ``standard'' cooling time at radius $r$.
(The cooling time calculated at radius $r$, is not exactly
the same as the average cooling time of the gas inward to $r$.
However, it was noticed above that most of the radiative
cooling rate inward to $r$ occurs close to $r$.
I therefore neglect the difference between these two values.)
The ratio between $\tau_{\rm rad}(r)$ and $\tau_{\rm cool} (r)$
is given by
\begin{equation}
\frac {\tau_{\rm rad}(r)}{\tau_{\rm cool} (r)} =
\frac {L_{\rm rad} - L_{\rm heat} } {L_{\rm radS}}.
\label{eq:tau1}
\end{equation}
 Substituting for $L_{\rm heat}$ and $L_{\rm rad}$ from
equations (\ref{eq:lh}) and (\ref{eq:lr}), respectively,
and using $L_{\rm heatS} = L_{\rm radS}$, gives to first
order in $T_\delta$
\begin{equation}
\frac {\tau_{\rm rad}(r)}{\tau_{\rm cool} (r)} =
\frac {T_\delta}{T (r) } \left(
\frac{6+21 \beta}{4+10 \beta} +\frac {5}{2}
- \frac{6+21\beta}{8+10\beta} r -\frac {1}{\beta} r
- \frac {5}{2} r \right) .
\label{eq:tau2}
\end{equation}
 As mentioned above $\beta \sim 0.3$.
 Equation (\ref{eq:tau2}) gives then
\begin{equation}
\frac {\tau_{\rm rad}(r)}{\tau_{\rm cool} (r)} =
\frac {T_\delta}{T (r) } (4.2-8.52 r) =
\frac {\Delta T(r)}{T (r) } \left( \frac {4.2-8.52 r}{r-1} \right), 
\qquad {\rm for} \qquad \beta=0.2,
\label{eq:tau3}
\end{equation}
and
\begin{equation}
\frac {\tau_{\rm rad}(r)}{\tau_{\rm cool} (r)} =
\frac {T_\delta}{T (r) } (4.3-6.2 r)=
\frac {\Delta T (r)}{T (r) } \left( \frac {4.3-6.2 r}{r-1} \right), 
\qquad {\rm for} \qquad \beta=0.4.
\label{eq:tau4}
\end{equation}
Note that $\beta=0.4$ is the case of constant pressure.
{{{ Although this is not realistic, in the assumption made here of
neglecting the pressure variation, the treatment of this case,
where there is no pressure gradient, is accurate.
The last two equations show that there are no fundamental differences
between the instabilities in the case of constant pressure
and the case of a shallow pressure gradient appropriate for clusters. }}}

{{{{
Let us examine the nature of the instability.
In a case of no radiative cooling, or a slow radiative cooling,
the restoring mechansim of a temperature perturbation is the heat
conduction itself.
Namely, when a region gets cooler the temperature gradient gets
steeper, the energy flow via heat conduction from warmer to cooler
regions increases, restoring the cooling region back to the equilibrium
temperature.
What was showen above is that under the conditions which exist
in cooling flow clusters, when a large inner regions gets cooler,
hence denser, the energy lose rate via radiative cooling, i.e.,
the radiative power, increases more than the heating power
via the heat conduction does.
For local perturbations, the short scale means steep
temperature gradient, hence efficient heat conduction.
This is why ZN03 find their perturbation to be stable.
I instead consider large scale perturbations, where radiative
energy lose occurs in a large region, as appropriate for the
inner cooling flow region.
Such large perturbations are reasonable in the violent
environment near the cD galaxy in cooling flow clusters. 
The time evolution of the departure from the initial perturbative
state occurs on the cooling time scale or longer
(since some heat conduction may exist).
Therefore, it is important (during the cluster age) only in the
inner regions of cooling flow clusters (and only in cooling flow
cluster) where cooling time is very short.
}}}}                                                             

For an instability to develop, the expression inside the parenthesis
in the far right-hand side of the last two equations should be
negative.
This occurs for $r>1$. However, for a large scale perturbation
this occurs at relatively large radii, where the standard cooling time
$\tau_{\rm rad}$ is long. 
There is also an instability at much smaller radii.
For these, the perturbation, as chosen here, must be a large scale
perturbation: It starts at $r=1$ (eq. \ref{eq:Tprime}), and the
instability occurs only for
$r<r_i=4.2/8.52 = 0.5$ for $\beta=0.2$, and
$r<r_i=4.3/6.2= 0.7$ for $\beta=0.4$.
I consider the case $\beta=0.2$ (the results are not sensitive to the
value of $\beta$), at $r=0.25$.
For example, the perturbation starts at $r=1$, which
in real units I take at $R=60~{\rm kpc}$, and I consider the region
around $r \simeq 0.25$, or in real units $R \simeq 10-20~{\rm kpc}$.
The ``standard'' cooling time in this region in cooling flow clusters
is $\tau_{\rm rad} (R=10-20~{\rm kpc}) \simeq 2-5 \times 10^8~{\rm yr}$.
Hence the cooling time, scaled with a perturbation of $\sim 10 \%$ is
\begin{equation}      
{\tau_{\rm cool}(15~{\rm kpc})} \simeq 2 \times 10^9
\left( \frac {\tau_{\rm rad}(15~{\rm kpc})}{5 \times 10^8~{\rm yr}}
\right)     \left( \frac {\vert \Delta T(15~{\rm kpc}) \vert}
{0.1 T (15~{\rm kpc})) } \right)
~{\rm yr},
\label{eq:tau5}
\end{equation}
for a perturbation given by equation (\ref{eq:Tprime}) with
$r=1$ at $R\sim 40-80~{\rm kpc}$.
For these parameters, the perturbation will grow to the nonlinear regime
during the age of the cluster, $\sim 10^{10} ~{\rm yr}$.
Therefore, a model with no cooling flow requires, either that
(1) the cooling flow is much younger than the cluster, e.g.,
there is an episodic heating of the cooling flow region
on a time scale of $\sim 10^9~{\rm yr}$
(e.g., Soker 2001), or
(2) there is another continuous energy source, e.g., an AGN at
the center of the central cD galaxy (Ruszkowski \& Begelman 2002).
In either cases, heat conduction by itself cannot ``kill'' the
cooling flow.

{{{ The following should be noted.
The instability occurs for $r <r_i$, whereas the region
$r_i < r < 1$ is stable.
However, this stable region cannot stabilize the unstable region
$r<r_i$.
This is because the instability evolves on a cooling time scale,
which is much shorter in the inner unstable region than in the
stable region.
The non-existence of a steady state solution was found by
Bregman \& David (1988) in their numerical calculations. }}}

\section{Summary}
New results by the Chandra and XMM-Newton X-ray telescopes,
which indicate a very low mass cooling rate at
temperatures bellow $\sim 1~{\rm kev}$ (see review by Fabian 2002),
renewed interest in heat conduction from the outer ICM as an
energy source to compensate for radiative cooling at lower
temperatures.
The goal of the present paper is to use a simple analytical
model to show that (simple) heat conduction models cannot
significantly suppress cluster cooling flows.
I used a power-law ICM and built a static medium where heat
conduction globally balances radiative cooling, and then perturbed it.
I showed that a large scale temperature perturbation with an
amplitude of $\sim 10 \%$, will grow to the non-linear regime
within a Hubble time.
The perturbation should extend over a large fraction of the
cooling flow region.
Such perturbations are reasonable in clusters which frequently
experience mergers and/or AGN activity, e.g., X-ray deficient
bubbles (McNamara 2002).
This result strengthens the finding of Bregman \& David (1988),
who use numerical calculations to show that a steady
solution does not exist for a constant heat conduction coefficient.

This instability over a time scale of $\sim 10^9 ~ {\rm yr}$,
implies that another energy source should either make the cooling
flow a short-lived episodic process, or that it continuously supplies
energy.
In either cases the heat conduction does not seem to play a
significant role; the major role is played by the
other energy source(s).
A popular energy source is AGN activity, which was proposed to
reduce the mass cooling rate in cooling flows in
galaxies (e.g., Binney \& Tabor 1995; Ciotti \& Ostriker 1997, 2001;
Jones et al. 2002), and in clusters of galaxies
(e.g., Soker et al. 2001; Churazov et al. 2002)

To the instability found in the present paper, I add the
failure of ZN03 to build a steady model with heat conduction
to five clusters,
{{{{ the claim made by Sun et al.\ (2003) that generally heat conduction
can't compensate for radiative cooling in cluster cooling flows, }}}}
the indications that heat conduction is suppressed at other regions
of clusters (Loeb 2002; Markevitch et al. 2002; Nath 2003),
and the required fine tuning argued by Bregman \& David (1988),
and conclude that heat conduction cannot explain the properties of
cooling flow clusters.

\acknowledgements
{{{{I thank the referee and Larry David for useful
comments. }}}
This research was supported in part by grants from the
US-Israel Binational Science Foundation and
the Israel Science Foundation.

\end{document}